 \documentclass[final,3p,times,twocolumn]{elsarticle}

 \usepackage{graphicx}

\usepackage{amssymb}
\usepackage{units}
\usepackage{url}

\journal{Physics Letters B}

\begin{document}

\begin{frontmatter}



\title{Experimental constraints on the $\gamma$-ray strength function in $^{90}$Zr using partial cross sections of the $^{89}$Y(p,$\gamma$)$^{90}$Zr reaction}


\author[col]{L.~Netterdon\corref{cor1}}
\ead{lnetterdon@ikp.uni-koeln.de}
\author[fra]{A.~Endres}
\author[ulb]{S.~Goriely}
\author[col]{J.~Mayer}
\author[col]{P.~Scholz}
\author[col]{M.~Spieker}
\author[col]{A.~Zilges}

\address[col]{Institute for Nuclear Physics, University of Cologne, Z\"ulpicher Stra{\ss}e 77, D-50937 Cologne, Germany}
\address[fra]{Institute for Applied Physics, Goethe University Frankfurt am Main, D-60438 Frankfurt am Main, Germany}
\address[ulb]{Institut d'Astronomie et d'Astrophysique, CP-226, Universit\'{e} Libre de Bruxelles, 1050 Brussels, Belgium}

\begin{abstract}
Partial cross sections of the $^{89}$Y(p,$\gamma$)$^{90}$Zr reaction have been measured to investigate the $\gamma$-ray strength function in the neutron-magic nucleus $^{90}$Zr. For five proton energies between \unit[$E_p=3.65$]{MeV} and \unit[$E_p=4.70$]{MeV}, partial cross sections for the population of seven discrete states in $^{90}$Zr have been determined by means of in-beam $\gamma$-ray spectroscopy. Since these $\gamma$-ray transitions are dominantly of $E1$ character, the present measurement allows an access to the low-lying dipole strength in $^{90}$Zr. A $\gamma$-ray strength function based on the experimental data could be extracted, which is used to describe the total and partial cross sections of this reaction by Hauser-Feshbach calculations successfully. Significant differences with respect to previously measured strength functions from photoabsorption data point towards deviations from the Brink-Axel hypothesis relating the photo-excitation and de-excitation strength functions.
\end{abstract}


\end{frontmatter}

\section{Introduction}
The $\gamma$-ray strength function, especially the $E1$-strength distribution, plays an important role as an ingredient in statistical-model calculations for nuclear astrophysics. This is the case for the synthesis of trans-iron elements during the so-called $p$ process \cite{Arnould03,Rauscher13}, as well as for neutron-capture reactions within the so-called $s$ and $r$ processes of nucleosynthesis \cite{Kaeppeler11,Arnould07}, which are called for to explain the 
origin of almost all nuclei  heavier than iron in the universe.

The low-lying electric dipole strength below the neutron separation energy has attracted a lot of attention recently, in particular due to the possible presence of a Pygmy Dipole Resonance (PDR)~\cite{Savran13}. The Nuclear Resonance Fluorescence (NRF) method is a very successful experimental technique to study dipole states selectively, mostly using electron bremsstrahlung for ($\gamma$,$\gamma$') measurements, see Ref.~\cite{Savran13} and references therein. However, also mono-energetic $\gamma$-rays obtained from laser Compton back-scattering were frequently used, see, $e.g.$, Ref.~\cite{Tonchev10}. Moreover, different experimental approaches using charged particles have been used to investigate the PDR; these include ($\alpha$,$\alpha$'$\gamma$) \cite{Poelhekken92,Endres10,Derya14} or (p,p') \cite{Poltoratska12,Iwamoto12} experiments. In addition, the Oslo method \cite{Schiller00} using ($^3$He,$^3$He'$\gamma$) and ($^3$He,$\alpha \gamma$) reactions was used for the determination of $\gamma$-ray strength functions at low $\gamma$-ray energies, for example in Cd or Pd isotopes \cite{Larsen13,Eriksen14}. 

In this Letter, we report on the investigation of the $\gamma$-ray strength function below and close above the neutron separation energy of \unit[$S_n=11968(3)$]{keV} \cite{AME2012}. Partial cross sections of selected $\gamma$-ray transitions of the $^{89}$Y(p,$\gamma$)$^{90}$Zr reaction are used to experimentally constrain the $\gamma$-ray strength function in the compound nucleus $^{90}$Zr at $\gamma$-ray energies between \unit[$E_\gamma = 7.71$]{MeV} and \unit[$E_\gamma = 12.98$]{MeV}. Partial cross sections were investigated for the $^{74}$Ge(p,$\gamma$)$^{75}$As reaction before \cite{Sauerwein12}, but no adjustment of the $\gamma$-ray strength function was necessary in this case.

\section{Experimental method}

\begin{figure}[tb]
\centering
\includegraphics[width=\columnwidth]{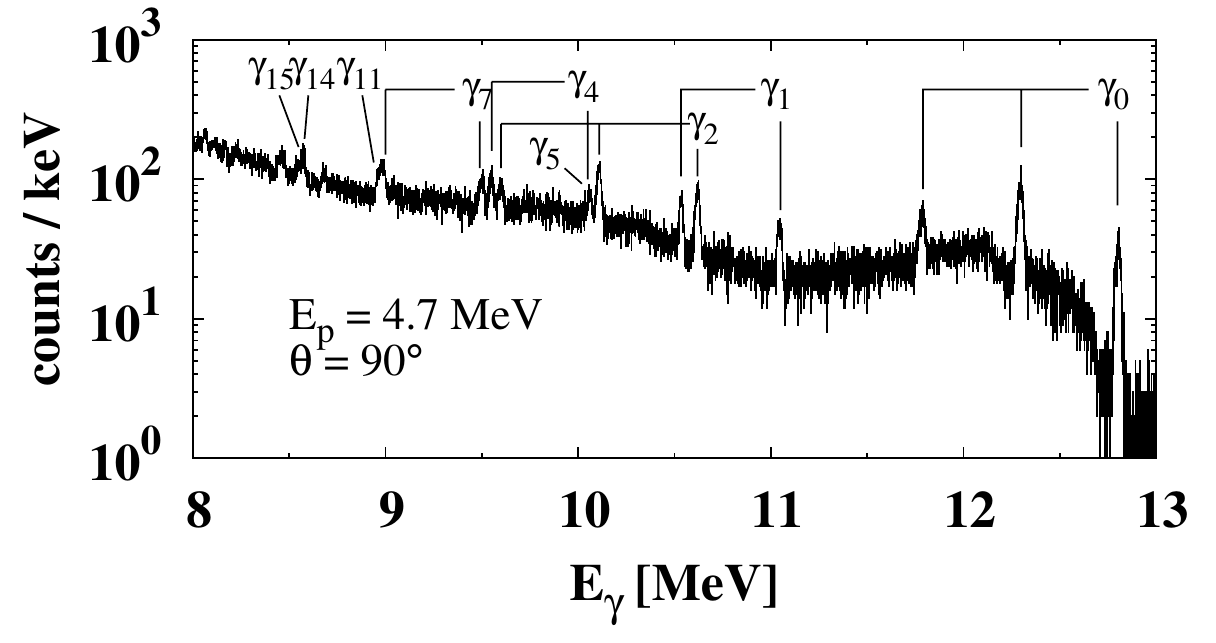}
\caption{High-energy part of a measured $\gamma$-ray spectrum using \unit[4.7]{MeV} protons. The spectrum was obtained by summing up the $\gamma$-ray spectra of all HPGe detectors placed at an angle of \unit[90]{$^\circ$} relative to the beam axis. The transition to the ground state is denoted as $\gamma_0$, to the first excited state as $\gamma_1$, and so on. De-excitations of the compound state up to the 15$^\mathrm{th}$ excited state in $^{90}$Zr are clearly visible. The single- and double escape peaks are marked as well, if visible. The excitation energies are adopted from Ref.~\cite{NNDC}.}
\label{fig:spectrum}
\end{figure}

The proton beam was delivered by the \unit[10]{MV} FN tandem ion accelerator at the Institute for Nuclear Physics at the University of Cologne, Germany. The beam with currents ranging from \unit[1]{nA} to \unit[60]{nA} impinged on a monoisotopic $^{89}$Y target with a thickness of \unit[583(24)]{$\mu$g/cm$^2$}, that was prepared by vacuum evaporation on a thick tantalum backing serving as a beam dump. The large range of the beam current was due to technical limitations of the tandem accelerator. The $^{89}$Y(p,$\gamma$)$^{90}$Zr reaction was studied by means of in-beam $\gamma$-ray spectroscopy at five different proton energies from \unit[$E_p=3.65$]{MeV} to \unit[$E_p=4.70$]{MeV} using the high-efficiency high-purity germanium (HPGe) detector array HORUS \cite{Netterdon14} . The $Q$ value of this reaction, which equals the proton-separation energy $S_p$, amounts to \unit[$Q=S_p=8353.4(1.6)$]{keV} \cite{AME2012}. The $\gamma$-ray spectrometer HORUS consists of up to 14 HPGe detectors, where six of them can be equipped with BGO shields for an active suppression of the Compton-scattered $\gamma$-rays. The detectors are placed at five different angles with respect to the beam axis in order to measure angular distributions of the $\gamma$-ray transitions, which were used to obtain absolute cross section values. Experimental details and the determination of total cross section values were already published in Ref.~\cite{Netterdon14}. In this Letter, the partial cross sections are discussed. The high-energy part of a typical $\gamma$-ray spectrum measured using \unit[4.7]{MeV} protons is shown in Figure~\ref{fig:spectrum}. This spectrum was obtained by summing the $\gamma$-ray spectra of the six HPGe detectors placed at an angle of \unit[90]{$^\circ$} relative to the beam axis. One can clearly identify de-excitations of the compound state up to the 15$^\mathrm{th}$ excited state in $^{90}$Zr. However, the transitions $\gamma_4$ and $\gamma_5$ could not be clearly separated in the spectra which hampered a reliable analysis of these partial cross sections. It should be noted, that a large number of unresolvable resonances with a width of the order of a few hundred eV are excited in the compound state instead of a single excited state due to the very high level density. Thus, the width of the peaks in Figure~\ref{fig:spectrum} is composed of the energy straggling inside the target material and the energy resolution of the HPGe detectors and amounts to approximately \unit[15]{keV} to \unit[30]{keV}, depending on the $\gamma$-ray energy. Hence, the observed $\gamma$-ray transitions are essentially a large number of transitions which cannot be separated by the HPGe detectors.

\begin{figure}[tb]
\centering
\includegraphics[width=\columnwidth]{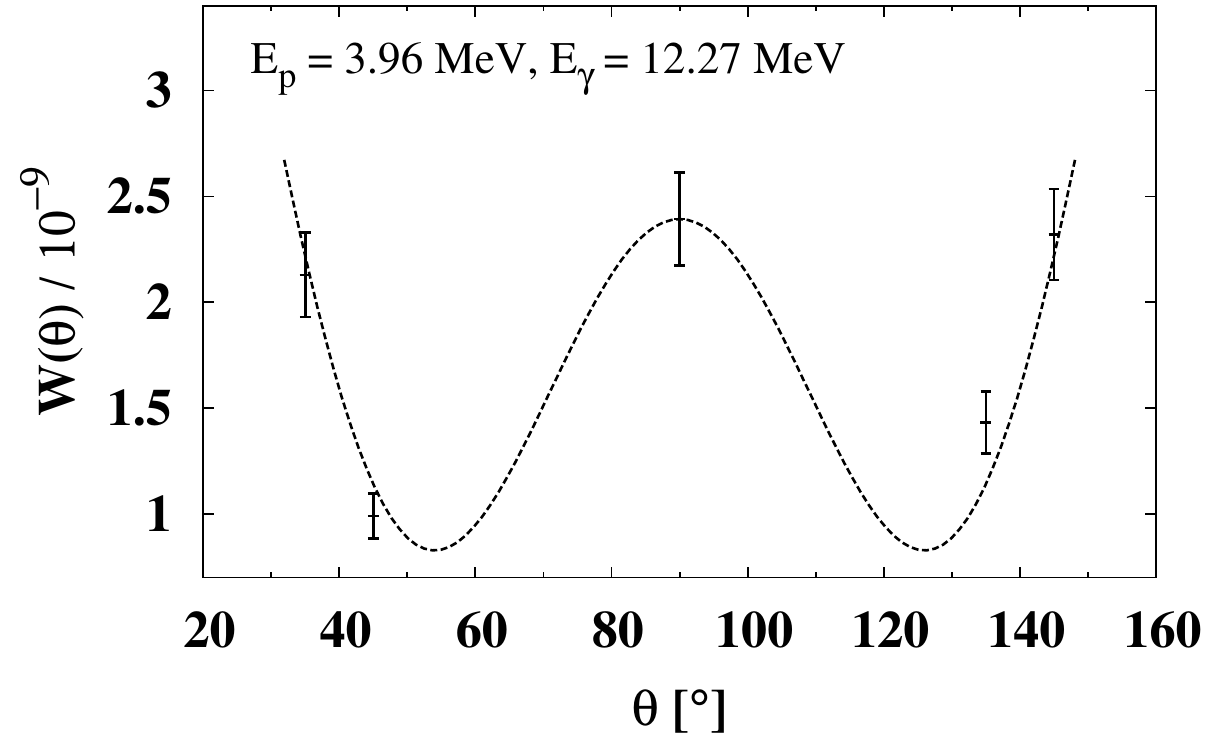}
\caption{Angular distribution for the $\gamma_0$ transition with an energy of \unit[$E_\gamma=12.27$]{MeV} in $^{90}$Zr for an incident proton energy of \unit[$E_p=3.96$]{MeV}. The dashed line corresponds to the sum of Legendre polynomials fitted to the experimental $W(\theta)$ values.}
\label{fig:angulardistribution}
\end{figure}

In order to determine absolute partial cross sections, the angular distributions of these high-energetic $\gamma$-ray transitions were measured. The measured intensities of each $\gamma$-ray transition for each angle are corrected for the number of impinged projectiles, full-energy peak efficiency, and dead time of the data-acquisition system. Afterwards, a sum of Legendre polynomials is fitted to the experimental angular distributions:

\begin{equation}
W^i \left(\theta\right) = A_0^i \left( 1 + \sum_{k = 2,4} \alpha_k P_k \left(\cos \theta\right) \right) \, ,
\end{equation} 
with the energy-dependent coefficients $A_0$, $\alpha_2$, and $\alpha_4$. The partial cross section $\sigma(\gamma_i)$ for the transition to the $i^{\mathrm{th}}$ excited state is then determined by normalizing the coefficients $A_0^i$ to the number of target nuclei. An example of an angular distribution for the $\gamma_0$ transition, i.e. the transition from the highly excited entry state to the ground state for a proton energy of \unit[$E_p=3.96$]{MeV} is shown in Figure~\ref{fig:angulardistribution}. For this specific angular distribution, the coefficients are $A_0 = (1.99 \pm  0.11) \times 10^{-9}$, $\alpha_2 = (2.30 \pm 0.56)$, and $\alpha_4 = (3.54 \pm 0.47)$. In total, seven partial cross sections could be measured for each proton energy.

\section{Results and discussion}

The experimental partial cross sections to the seven lowest $0^+$ and $2^+$ levels are listed in Table~\ref{tab:results}. Since after the (p,$\gamma$) reaction dominantly $1^-$ states are populated, the electromagnetic de-excitation of the compound nucleus is dominated by the $E1$ mode. The experimental uncertainties are composed of the uncertainties in target thickness (\unit[$\approx 4$]{\%}), accumulated charge (\unit[$\approx 5$]{\%}), full-energy peak efficiency  (\unit[$\approx 8$]{\%}), and statistical errors (\unit[$\approx 4 - 7$]{\%}). The remaining $\gamma$-ray transitions were not visible in the spectra since they are strongly suppressed by electro-magnetic selection rules. For example, the third excited state has a spin and parity configuration of $J^\pi = 5^-$ \cite{NNDC}. 

\begin{figure}[!b]
\centering
\includegraphics[width=\columnwidth]{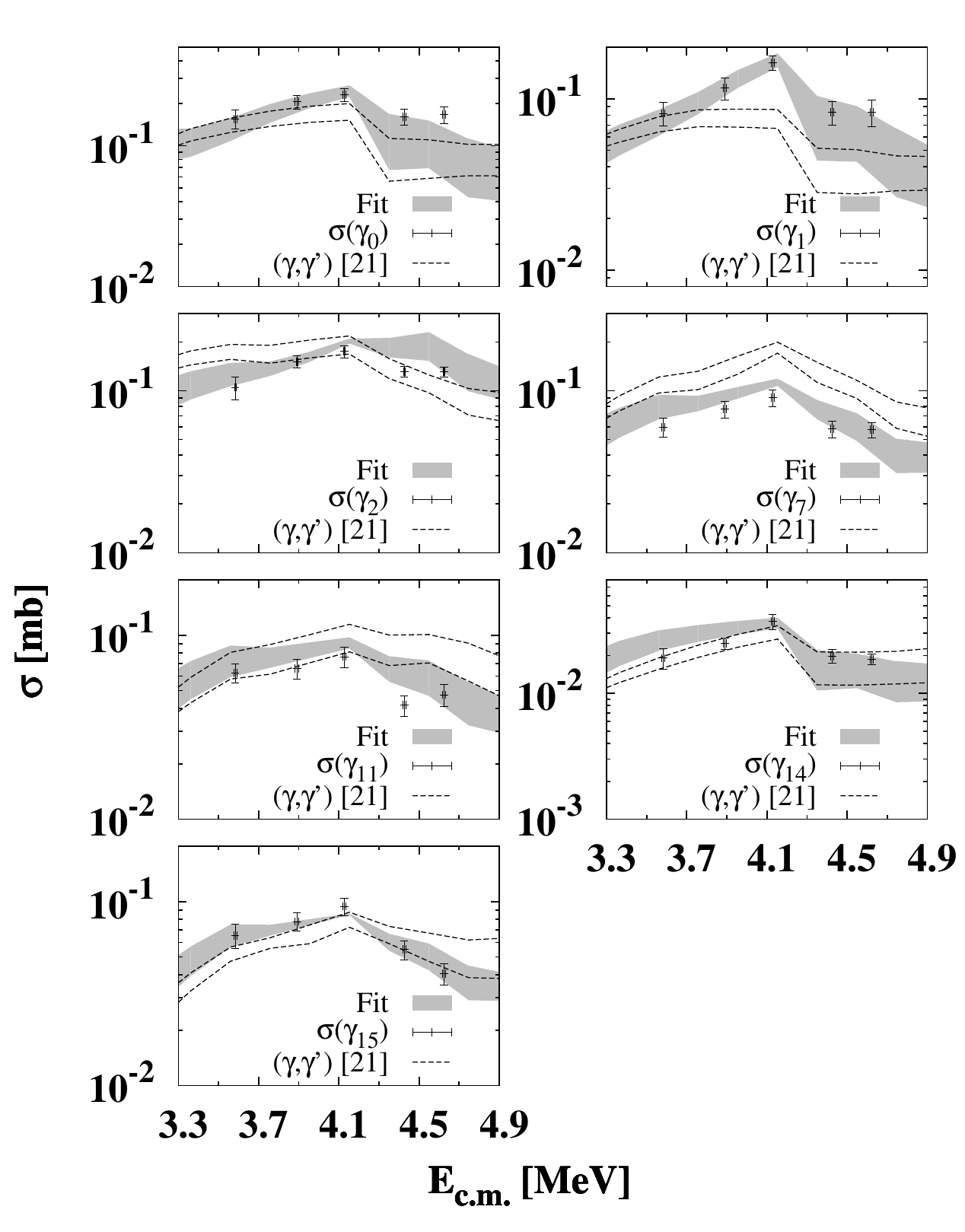}
\caption{Experimental partial cross sections of the $^{89}$Y(p,$\gamma$)$^{90}$Zr reaction. The data are compared to \textsc{TALYS} calculations using the $\gamma$-ray strength function extracted from the ($\gamma,\gamma$') measurement of Ref.~\cite{Schwengner08}. The dashed lines correspond to the experimental uncertainty of the ($\gamma,\gamma$') data. The gray shaded areas show a \textsc{TALYS} calculation using a $\gamma$-ray strength function fitted to the experimental partial cross sections. These areas include the uncertainties related to the proton-nucleus OMP, NLD, and $M1$ strength in $^{90}$Zr.}
\label{fig:results}
\end{figure}

In Figure~\ref{fig:results}, the experimental data are compared to calculations using the statistical model code \textsc{TALYS} \cite{Talys,Talys12}. First of all, it was found, that no available standard theoretical $\gamma$-ray strength function was capable of correctly describing all experimental partial cross sections consistently. It should be stressed that the $E1$ strength function has recently been extracted from a ($\gamma$,$\gamma$') measurement \cite{Schwengner08}. Based on the Brink-Axel hypothesis, this $\gamma$-ray strength function should lead to a correct description of the partial cross sections, when it is used as an input for \textsc{TALYS}. However, comparing the \textsc{TALYS} calculation using this $E1$ strength with the experimental data reveals discrepancies that cannot be attributed to other nuclear ingredients in the model. 
The partial cross sections calculated using the $\gamma$-ray strength function deduced from the ($\gamma$,$\gamma'$) experiment and the one fitted to the present experimental data show in general a similar behavior. However, noticeable differences appear for the $\gamma_1$ transition around \unit[$E_\mathrm{c.m.} \approx 4$]{MeV}, the $\gamma_7$ transition, as well as for the $\gamma_{11}$ transition above approximately \unit[4.5]{MeV}.
In an earlier measurement, the excitation curve of the $^{89}$Y(p,$\gamma_0$) reaction, i.e. the emission of the $\gamma_0$ transition, has been measured \cite{Obst66}. The general energy dependence is consistent with the present measurement of the partial cross-section of the $\gamma_0$ transition, especially at higher energies after the (p,n) channel has opened.

\begin{table*}[!ht]
\centering
\caption{Excitation energy ($E_x$), spin ($J$) and parity ($\pi$) of the seven levels of interest in the present measurements, as well as experimental partial cross sections $\sigma$($\gamma_i$) for the $^{89}$Y(p,$\gamma$) reaction as a function of the effective center-of-mass energy $E_\mathrm{c.m.}$. Partial cross sections are given in mb.}
\vskip 0.2cm
\begin{tabular}{ccccccccc}
\hline
$E_x$ [MeV] & & 0 & 1.7607 & 2.1863 & 3.3088 & 3.8422 & 4.1260 & 4.2230 \\
$J^{\pi}$ & & $0^+$ & $0^+$ & $2^+$ & $2^+$ & $2^+$ & $0^+$ & $2^+$ \\ 
\hline
$E_\mathrm{c.m.}$ [keV] & $E(\gamma_0)$ [keV] & $\sigma$($\gamma_0$)  &  $\sigma$($\gamma_1$) &  $\sigma$($\gamma_2$) & $\sigma$($\gamma_7$) &  $\sigma$($\gamma_{11}$)  & $\sigma$($\gamma_{14}$) & $\sigma$($\gamma_{15}$)\\ \hline
\vspace{-0.2cm}
 & & & & & & & & \\
3583(8)	& 11936(26)	& 0.16(2)  & 0.08(1) & 0.10(2) & 0.060(6) & 0.062(8) & 0.019(4) & 0.065(10) \\
3891(8)	& 12244(26)	& 0.20(2)  & 0.12(2) & 0.15(1) & 0.077(7) & 0.066(8) & 0.025(3) & 0.078(9)\\
4129(8)	& 12482(28)	& 0.23(2)  & 0.16(2) & 0.18(1) & 0.090(11)& 0.076(10) & 0.037(5) & 0.094(10) \\
4426(8)	& 12779(29)	& 0.16(2)  & 0.08(1) & 0.13(1) & 0.058(9) & 0.042(5) & 0.020(2) & 0.055(7)  \\
4624(8)	& 12977(26)	& 0.17(2)  & 0.08(2) & 0.13(1) & 0.057(8) & 0.047(7) & 0.019(2) & 0.040(5) \\ \hline
\end{tabular}
\label{tab:results}
\end{table*}

\begin{figure}[tb]
\centering
\includegraphics[width=\columnwidth]{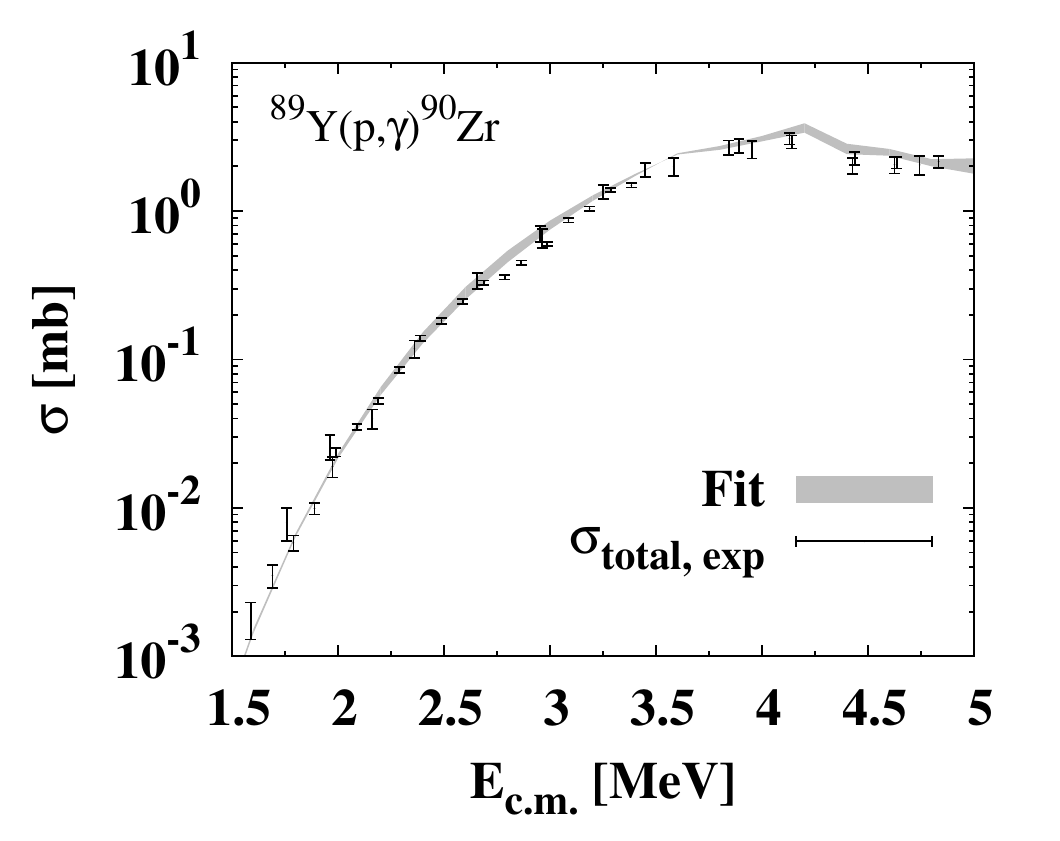}
\caption{Total cross section values of the $^{89}$Y(p,$\gamma$)$^{90}$Zr reaction. The experimental data from Refs.~\cite{Netterdon14,Harissopulos13} are compared to \textsc{TALYS} calculations using the adjusted $\gamma$-ray strength function including the theoretical uncertainty stemming from the proton-OMP, nuclear level density, and $M1$ strength. Within these uncertainties, an excellent agreement is found over the whole energy region.}
\label{fig:total}
\end{figure}

The $E1$ strength function was adjusted independently in order to reproduce the total and partial cross sections at best, simultaneously for the seven channels. The experimental total cross section~\cite{Netterdon14,Harissopulos13} is shown in Figure~\ref{fig:total} and compared  with \textsc{TALYS} calculations using the adjusted $\gamma$-ray strength function, as discussed below.

The adjustment of the $\gamma$-ray strength function was only possible within the $\gamma$-ray energy range, which is accessible through the measured partial cross sections, i.e. from \unit[$E_\gamma = 7.71$]{MeV} to \unit[$E_\gamma = 12.98$]{MeV}. Below and above this energy range, the microscopic D1M+QRPA $E1$ strength  \cite{Martini14} based on the finite-range D1M Gogny interaction \cite{Goriely09} was used, since it reproduces rather well the giant dipole resonance region as experimentally constrained by photoneutron data \cite{Berman67,Lepretre71}. The calculated partial cross sections using the adjusted $E1$-strength are shown as gray-shaded areas in Figure~\ref{fig:results}. The shaded areas reflect the uncertainty analysis with respect to other ingredients of the Hauser-Feshbach calculation, namely the proton-nucleus optical model potential (OMP), the nuclear level density (NLD), and $M1$ strength function. As far as the OMP is concerned, both the semi-microscopic proton-OMP of Ref.~\cite{Bauge01} and the phenomenological proton-OMP of Ref.~\cite{Koning03} are considered and yield a good description of the total cross section (see Figure~\ref{fig:total}). Both were used to calculate the partial cross sections and the differences are taken into account in the uncertainty of the calculations. Various models for the NLD lead to very similar partial cross sections. Essentially, the total cross section is affected by different models of the NLD above the (p,n) threshold. The combinatorial NLD using a temperature-dependent HFB approach of Ref.~\cite{Hilaire12} is seen in Figure~\ref{fig:total} to give an excellent agreement with the total cross section above the (p,n) threshold, and is consequently used to calculate the partial cross sections. Finally, the $M1$ strength is known to be rather strong in $^{90}$Zr, as revealed by an inelastic proton-scattering experiment \cite{Iwamoto12}. The $M1$ resonance is found to be located at an energy of \unit[$E_{M1}=9.53(6)$]{MeV} with a width of \unit[$\Gamma_{M1}=2.70(17)$]{MeV}. In a recent work, the fine structure of the giant $M1$ resonance was investigated by means of a photon-scattering experiment \cite{Rusev13}, which revealed a sum strength of \unit[4.17(56)]{$\mu_N^2$}. This $M1$ resonance contribution, in turn, also influences the adjustment of the $E1$ strength. The experimentally obtained $M1$ strength was included in the \textsc{TALYS} calculation and its experimental uncertainties were included in the determination of the shaded areas shown in Figure~\ref{fig:results}. The same uncertainty analysis using the $E1$-strength of Ref.~\cite{Schwengner08} was performed and is shown by the dashed lines in Figure~\ref{fig:results}.

\begin{figure}[tb]
\centering
\includegraphics[width=\columnwidth]{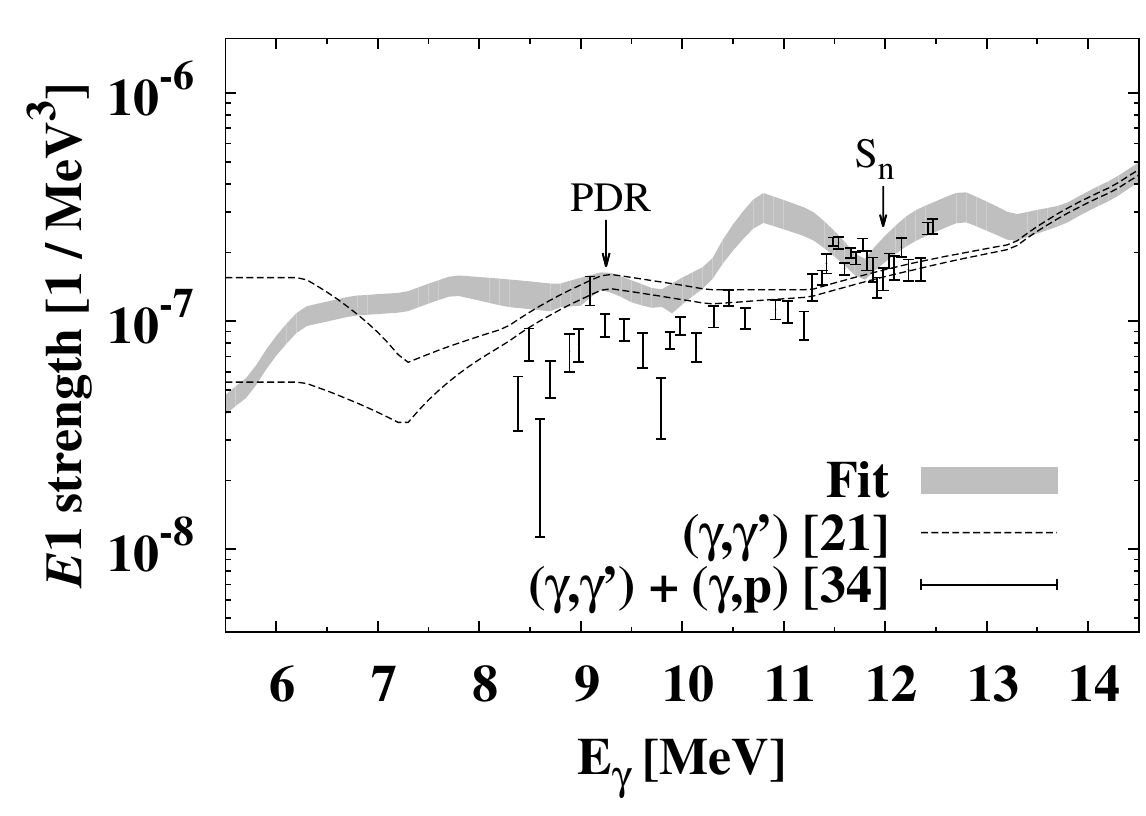}
\caption{$E1$-strength distribution as a function of $\gamma$-ray energy in $^{90}$Zr extracted from the measurement of partial cross sections of the $^{89}$Y(p,$\gamma$)$^{90}$Zr reaction. The gray shaded area depicts the adjusted $\gamma$-ray strength function used to reproduce the experimental partial cross sections by Hauser-Feshbach calculations. Additionally shown is the $\gamma$-ray strength function obtained from a ($\gamma,\gamma$') measurement of Ref.~\cite{Schwengner08} and of Ref.~\cite{Axel70}. The dashed lines correspond to the experimental uncertainty of the ($\gamma,\gamma$') data. The strength around the PDR energy of about 9~MeV as found in Refs.~\cite{Iwamoto12,Schwengner08} is well reproduced by the present measurement, but a significant enhancement is observed around the neutron-separation energy $S_n$.}
\label{fig:gammastrength}
\end{figure}

The final $E1$-strength distribution extracted to reproduce the experimental partial cross sections is shown in Figure~\ref{fig:gammastrength}. First, it is clear that the PDR measured in Ref.~\cite{Schwengner08} at an energy of \unit[$E_\mathrm{PDR} \approx 9$]{MeV} with a width of \unit[$\Gamma_\mathrm{PDR} \approx 2$]{MeV} is in excellent agreement with our present result. However, at lower and higher $\gamma$-ray energies, both $\gamma$-ray strength functions differ significantly. A possible reason for this disagreement might be the $\gamma$-branching ratio of the PDR states. In Ref.~\cite{Schwengner08}, an average branching ratio was obtained from a statistical-model calculation. However, it was found for the close-lying nucleus $^{94}$Mo, that calculated and experimental branching ratios can differ by approximately a factor of two \cite{Romig13}. In contrast, it is worth mentioning here, that mean branching ratios were found to be consistent with experimentally determined ones \cite{Massarczyk14}.  Nevertheless, partial cross section data present the advantage of being independent from any prediction of the branching ratios. However, the branching ratios cannot account for the difference at higher energies around the neutron separation energy of \unit[11968(3)]{keV} \cite{AME2012}. A strong enhancement of the $E1$-strength around \unit[$E_\gamma \approx 11$]{MeV} and \unit[$E_\gamma \approx 13$]{MeV} is inevitable to reproduce the measured partial cross sections to the ground and first excited states, despite the fact, that the cross-section drops due to the opening of the (p,n) channel. Naturally, photoabsorption cross sections can only be deduced from ($\gamma,\gamma$') experiments up to the neutron separation energy. Thus, the enhancement of the $E1$ strength around \unit[$E_\gamma \approx 13$]{MeV} could not have been observed in Ref.~\cite{Schwengner08}. Above $S_n$, the $E1$ strength in Ref.~\cite{Schwengner08} was obtained by combining \textsc{TALYS} calculations for the ($\gamma$,p) channel and experimental ($\gamma$,n) data from Ref.~\cite{Berman67}. 

Figure~\ref{fig:gammastrength} additionally shows the $\gamma$-ray strength function obtained from the photoabsorption experiment of Ref.~\cite{Axel70}. Compared to the ($\gamma$,$\gamma$') data of Ref.~\cite{Schwengner08}, a larger strength is observed at around \unit[$E_\gamma = 12.5$]{MeV}, but not as significant as in the present experiment. A similar picture arises for the enhancement of the $E1$ strength at about \unit[$E_\gamma = 11.5$]{MeV}. However, this $\gamma$-ray strength function differs from the one obtained in the present work, especially regarding the strong enhancement around \unit[$E_\gamma \approx 11$]{MeV} and \unit[$E_\gamma \approx 13$]{MeV}. In the present measurement, no partial cross-sections for $\gamma$-ray energies around \unit[12]{MeV} could be measured. Thus, it is not possible to exclude a stronger enhancement around \unit[$E_\gamma \approx 12$]{MeV} from the present experimental data.
It should be stressed in this context, that both $E1$-strength distributions \cite{Schwengner08,Axel70} are deduced from photo-excitation measurements, whereas the one fitted to the partial cross-sections is based on the photon de-excitation of the compound nucleus. The generally larger strength and enhanced low-energy tail deduced from the present experiment might hint towards some deviation with respect to the Brink-Axel hypothesis that assumes that the de-excitation strength function from excited states can be directly associated with the excitation strength function from the ground state. The resulting temperature dependence of the $\gamma$-ray strength function can hardly be confirmed from the present experimental results since the \unit[$E_\gamma \approx 13$]{MeV} pattern is exclusively based on the $\gamma_0$ transitions and the \unit[$E_\gamma \approx 11$]{MeV} peak on the $\gamma_1$ transitions without any overlap between both cases. This result also suggests that the presently found $\gamma$-ray strength function should be used for the de-excitation channel.

\begin{table*}[!ht]
\centering
\caption{Stellar reaction rates for the $^{90}$Zr($\gamma$,n)$^{89}$Y reaction for typical $p$-process temperatures in cm$^{3}$ s$^{-1}$ mole$^{-1}$. Reaction rates were calculated using the presently obtained $\gamma$-ray strength function (Fit), as well as the one from Ref.~\cite{Schwengner08}. These are additionally compared to results from the \textsc{NonSmoker} code \cite{Nonsmoker} and the \textsc{BRUSLIB} library \cite{Goriely08, Xu2013}. Large deviations are found for different adopted models of the $E1$-strength distribution.}
\vskip 0.2cm
\begin{tabular}{ccccc}
\hline
T [GK]  & Fit & ($\gamma,\gamma'$) \cite{Schwengner08} & \textsc{NonSmoker} & \textsc{BRUSLIB} \\
\hline
2.0		& 1.46 $\times$ 10$^{-11}$	&	1.26 $\times$ 10$^{-11}$ & 4.14 $\times$ 10$^{-11}$	& 6.80 $\times$ 10$^{-12}$\\
2.5	& 2.18 $\times$ 10$^{-5}$		&   1.87 $\times$ 10$^{-5}$ & 6.48 $\times$ 10$^{-5}$	&	1.04 $\times$ 10$^{-5}$		 \\
3.0			& 3.00 $\times$ 10$^{-1}$		&   2.56 $\times$ 10$^{-1}$ & 9.20 $\times$ 10$^{-1}$	&	1.47 $\times$ 10$^{-1}$\\
\hline
\end{tabular}
\label{tab:reactionrate}
\end{table*}

In order to illustrate some astrophysical impact of the experimentally determined $E1$ strength, the stellar $^{90}$Zr($\gamma$,n)$^{89}$Y reaction rate for typical $p$-process temperatures of \unit[2 - 3]{GK} was calculated using the $\gamma$-ray strength function obtained from the present experiment, as well as the one of Ref.~\cite{Schwengner08}. In addition, these calculations were compared to reaction rates obtained with the \textsc{NonSmoker} code \cite{Nonsmoker}, as well as the \textsc{BRUSLIB} library \cite{Goriely08, Xu2013}, see Table~\ref{tab:reactionrate}. For a temperature of \unit[2]{GK}, the stellar reaction rate using the presently obtained $E1$-strength is higher by about \unit[15]{\%} than the rate using the $E1$ strength of Ref.~\cite{Schwengner08}. In contrast, the stellar rate is enhanced by a factor of two compared to the \textsc{BRUSLIB} results based on the QRPA strength \cite{Goriely2004} and decreased by about a factor of three compared to the \textsc{NonSmoker} results based on a Lorentzian-type E1 strength function \cite{Cowan91}. In addition, the temperature dependence of the stellar reaction rates differs significantly for different models of the $\gamma$-ray strength function. Since the $\gamma$-ray strength function is one of the main ingredients for the calculation of stellar reaction rates, these results underline the importance of a correct understanding and description of the $\gamma$-ray strength function.


\section{Summary and conclusions}
In summary, partial cross sections of the $^{89}$Y(p,$\gamma$)$^{90}$Zr reaction were measured by means of in-beam $\gamma$-ray spectroscopy. Seven partial cross section for each of the five proton energies could be obtained and are used for the first time to determine indirectly the $E1$ strength function. These data  experimentally constrain the low-lying dipole strength of $^{90}$Zr at $\gamma$-ray energies between \unit[$E_\gamma = 7.71$]{MeV} and \unit[$E_\gamma = 12.98$]{MeV}. The PDR around \unit[$E_\gamma \approx 9$]{MeV} that was already measured via the NRF method and inelastic proton-scattering is well reproduced by the present measurement. In contrast, a strong enhancement of the $E1$ strength around the neutron-separation energy is found. Using the $E1$-strength distribution from an earlier ($\gamma,\gamma$') measurement and the presently fitted one as an input for \textsc{TALYS} calculations leads to partly incompatible results of the partial cross sections. These differences may reflect deviations from the Brink-Axel hypothesis relating the photo-excitation and de-excitation strength functions. The stellar reaction rate was calculated using the fitted $\gamma$-ray strength function as well as the one obtained from ($\gamma,\gamma'$) data. The results were also compared to results from the \textsc{NonSmoker} code and the \textsc{BRUSLIB} library. Large deviations were found in the stellar reaction rates, when different models of the $\gamma$-ray strength function are applied. In the relevant temperature region for $p$-process nucleosynthesis, the new stellar reaction rate differs by factors from two to three to the presently used ones. This underlines the importance of a detailed knowledge of the $\gamma$-ray strength function for astrophysical applications. The measurement of partial cross section is a powerful experimental tool to investigate the low-lying $E1$-strength distribution regarding the nuclear structure of the involved nuclei as an important ingredient for Hauser-Feshbach calculations in nuclear astrophysics.

\section*{Acknowledgements}
The authors thank A. Dewald and the accelerator staff at the Institute for Nuclear Physics in Cologne for providing excellent beams. Moreover, we gratefully acknowledge K.~O. Zell for the target production, H.-W. Becker and D. Rogalla of the Ruhr-Universit\"at Bochum for the assistance during RBS measurements. This project was partly supported by the Deutsche Forschungsgemeinschaft under contract DFG(ZI 510/5-1), the ULDETIS project within the UoC Excellence Initiative institutional strategy, and the HIC for FAIR within the framework of LOEWE launched by the state of Hesse, Germany. S.G. is F.N.R.S. research associate.












\end{document}